\documentstyle[12pt,epsfig]{article}
\begin{document} 

\bigskip
\centerline{{\em A Ciencia Cierta} {\bf 2}(12)}
\centerline{\em \scriptsize The Bimestral Publication of the Potosinian Commettee of Science \& Technology}
\centerline{\em San Luis Potos\'{\i}}
\centerline{August 2002}

\bigskip

\centerline{Enlarged version in English}
\centerline{September 2002}

\bigskip
\bigskip

\centerline{\Large \bf ARISTOTLE: THE FIRST ENCYCLOPEDIST}

\bigskip

\centerline{Haret C. Rosu\footnote{e-mail: hcr@ipicyt.edu.mx; IPICyT, San Luis Potos\'{\i}, Mexico}}

\bigskip

\noindent
{\scriptsize {\bf Abstract.} Aristotle was the first to declare himself in the class of  natural philosophers ({\em physiologoi}) and he was the first 
physicist because he spelled out the first definitions of some basic physical concepts. Moreover, he was the first encyclopedist
of the world and perhaps the greatest.}  

\bigskip

In the January 2002 issue of the IOP magazine {\em Physics World}, M. Rowan-Robinson published the paper 
``Was Aristotle the first physicist ?" \cite{1}. This paper generated several reactions \cite{2} and stimulated me to learn more
about one of the greatest thinkers of all times. I believe the answer to the question of Rowan-Robinson is positive,
first of all because the first known physics books that have been conserved through the centuries belong to Aristotle.
But Aristotle was much much more. He was the first encyclopedist of mankind. Because of his extremely imaginative mind he
left behind him a multitude of systematic studies on almost all of the cultural fields of his epoch, including cosmology.
He is the founder of Logics, where he developed the syllogism as a basic argument of the theory of deduction. The syllogism
starts from two different premises that lead to a unique conclusion. Aristotle revised many of Plato's ideas insisting on methods
based on observation and experiment (he was practically a very acute observer of nature whereas his `experiments' were mainly
dissections). Aristotle argued that mankind, though a component of nature, was different from the rest of the known world due to his capacity for
thinking and reasoning. In his opinion, human beings are rational animals and despite the fact they do not act always according
to reason, they have always the choice of exerting this superior quality. To be happy, human beings have the ethical duty to
live in the realm of reason. Concerning politics, he believed that a monarchy with a clever king could be an ideal social structure,
but he agreed that a limited democracy may be one of the best compromises. The arts have been interpreted by Aristotle as a form of
intelectual pleasure. His analyses of the Greek tragedies have been used as models of literary critical commentaries. Moreover, he asserted that 
the observed material objects as well as the persons examining them are under continuous transformations. But since knowledge should
be associated only with the universal unchangeable elements Aristotle claimed that knowledge and perception are fundamentally
different, one from the other. As for the basic blocks of the material world, Aristotle believed that in addition to the four elements of
Empedocles - fire, air, earth, and water - there was a fifth one that he called ether filling the whole universe and being actually the 
unique element in the composition of any body `beyond the Moon'. The reason to introduce the ether was Aristotle's dogma that nature
abhors empty space (vacuum).

It is worth mentioning that the lecture notes of most of his courses in the Lyceum were discovered by chance by Roman soldiers
about two centuries after the death of Aristotle in a pit on the west coast of Asia Minor. The peripathetic philosopher Andronicus provided 
a Latin translation and tried 
without much success to publish a first edition of Aristotle's works; the project remained merely at the level of 
collection. The publication at a larger scale was achieved by the Arabs much later.

The works of Aristotle became known in the early medieval Europe of the XIth-XIIth centuries by translation from the arabic and rapidly became
recognized textbooks by the Christian Church despite their laical origin. Having this religious support, Aristotle's doctrines dominated  
scientificly oriented people until the Renascence.

As far as Physics is concerned, the first drawback came with Galilei, when his theory of free fall was first questioned from the 
experimental point of view\footnote{It is to be noted that within nonrelativistic quantum mechanics free fall could be still considered in 
a certain sense to be aristotelian, i.e., mass-dependent.}. With respect to the nature of time, Aristotle was the proponent of a uniform and 
continuous {\em flux} of time, an idea supported later by Newton. Finally, the ether paradigm was a powerful concept for centuries till the 
beginning of the XXth century when Einstein formulated the theory of special relativity. However, one should say that the surprizes of 
technological progress do not forbid the resurgence of different forms of ether-like concepts at least as a convenient explicative 
necessity\footnote{As a matter of fact, vacuum quantum fluctuations can be considered as a form of quantum ether. In addition, the so-called 
quintessence, which is in vogue in the cosmology produced by the high redshifted supernovae data and also some of the modern
cosmological scalar fields can be viewed as ethereal fields.}. Aristotle was an antiatomist. He believed that matter was continuous and could be 
devided in ever smaller pieces forever! On the other hand, Aristotle was the first to provide definitions of basic mechanical concepts, such as
energy ({\em energeia} - activity) and dynamics ({\em dynamis} - capacity for action). He also first defined physics ({\em physis} -nature) as the science
of the forms of motion.      

 \bigskip 


\section*{Biographical Sketch of Aristotle}

\bigskip

\noindent
$\bullet$	Period of life: 384 - 322 BC (62 years).\\

\noindent
$\bullet$    Place of birth: Stagira in the ancient Macedon region\footnote{Ancient Macedon region and people should not be confused with present Macedonia.}, 
now in the prefecture of Khalkidhiki in the northeastern part of Greece).
                 Aristotle was born as the son of Nicomachus, chief physician at the court of Amyntas III.
                 Stagira was a port in the three-fingered hand-like peninsula Calcidia and is now a village of only 500 people.\\ 

\noindent
$\bullet$	At the age of 17 years he went to Athens to study in the Academy, the famous educational institution of Plato (now equivalent to a department of 
        philosophy). For the next 19 years he remained in Plato's Academy where he reached the mature status of professor and independent research philosopher.
          Although many of his ideas were opposed to those of Plato, he was considered as one of the possible successors by Plato himself.\\

\noindent
$\bullet$	In 347 BC, when Plato died, Aristotle could not stay as head of the Academy since Plato's last will was that his own nephiew, the much more obscure Speusippus, be in charge of it, but probably also because of the opposition of the other philosophers and his Macedonian origin. Together with Xenocrates, who later became the third head of Plato's Academy, Aristotle abandoned Athens spending the following 12 years in various cities of the isles and along the coasts of the Aegean Sea where  for a while he engaged in his remarkable studies in biology.  
Fortunately, during this quite unstable period of his life (although he got married to Pythias, the niece of ruler Hermias) he was asked by
Philip II of Macedonia to come back to Stagira as a private mentor of his thirteen-year old son Alexander (later, Alexander the Great).\\

\noindent
$\bullet$	In 335 BC Philip conquered Athens. Aristotle returned to the most important city of that epoch in a position allowing him to found his own school (known as
the ``peripatetic school" because of Aristotle's habit of walking during his lectures). The place he chose for his school was known as the Lyceum named so after the nearby temple of
Apollo Lyceus ({\em Lykaios}), i.e., Apollo ``the Wolf-God". Aristotle's school location was discovered only in 1997 by E. Ligouri. It is a quadratic place with a diagonal of about 50 meters. It is considered by many historians as the first university of the planet\footnote{I think this is true if one takes into account the encyclopedic range of 
Aristotle's courses and the extended library that was accumulated there through the efforts of Aristotle.}. Aristotle taught in the Lyceum during the next 12 years of his life. His courses 
(2000 pages published in `Complete Works' representing only between a fifth and a quarter of his total cultural output) cover knowledge ranging from Ethics to Biology, Physics and Metaphysics.

On the personal side, it is known that during his second Athenian period his wife Pythias died and subsequently he lived unmarried with Herpyllis, a woman from Stagira.
Also, in 327 BC, his nephiew Callisthenes who went with Alexander in his famous military campaign was executed at the order of Alexander !\\

\noindent
$\bullet$	In 323 BC, at the death of Alexander, Aristotle prudently abandoned Athens and died a year later in Calchis, hometown of his mother in Euboea island.

\bigskip

\section*{Extant Aristotelian Treatises}

Presented is the list as given by Turner \cite{catenc} with the titles in Latin as assigned by Andronicus.  

\begin{verse}

{\bf Logical Treatises (also known as Organon)}

1.  Categoriae\\

2.  De Interpretatione\\

3.  Analytica Priora\\

4.  Analytica Posteriora\\

5.  Topica\\

6.  De Sophisticis Elenchis\\

\bigskip

{\bf Physical Treatises}

7.  Physica (or Physica Auscultatio)

8.  De C\oe lo

9.  Meteorologica

\bigskip

{\bf Metaphysical Treatises}

10. Metaphysica\footnote{Title given by Andronicus, who placed this treatise after the physical treatises in his 
collection; Aristotle entitled it `First Philosophy' 
({\em Prote Philosophia}).}

\bigskip

{\bf Biological and Zoological Treatises}

11. Historiae Animalium

12. De Generatione et Corruptione

13. De Generazione Animalium

14. De Partibus Animalium

\bigskip

{\bf Psychological and Anthropological Treatises}

15. De Anima

16. De Sensu et Sensibili

17. De Memoria et Reminiscentia

18. De Vita et Morte

19. De Longitudine et Brevitate Vitae

\bigskip

{\bf Ethical and Political Treatises}

20. Ethica Nicomachea

21. Politica

\bigskip

{\bf Poetical and Rhetorical Treatises}

22. De Poetica

23. De Rhetorica

\end{verse}

\bigskip
\bigskip

\noindent
{\bf Figure Caption}: Map showing the location of Stagira among other ancient Greek cities.


\begin{thebibliography}{99}

\bibitem{1} M. Rowan-Robinson, ``Was Aristotle the first physicist ?", {\em Physics World} {\bf 15}, 15-16 (January 2002), http://physicsweb.org

\bibitem{2} Y. Ne'eman, ``Aristotle: top man", {\em Physics World} {\bf 15}, 18-19 (May 2002); D. Sathe, untitled, {\em ibidem}, p. 19

\bibitem{catenc} W. Turner, ``Aristotle", in {\em The Catholic Encyclopedia, Volume 1} (R. Appleton Company, 1907) 

\end{thebibliography}
\end{document}